# Banana for scale: Gauging trends in academic interest by normalising publication rates to common and innocuous keywords

Edwin S. Dalmaijer [1], Joram Van Rheede [2], Edwin V. Sperr [3], & Juliane Tkotz [4]

*1. MRC Cognition and Brain Sciences Unit, University of Cambridge*
*2. MRC Brain Network Dynamics Unit, University of Oxford*
*3. Augusta University / University of Georgia Medical Partnership*
*4. Department of Clinical Psychology, Central Institute for Mental Health (Mannheim)*

**Corresponding author**

Dr Edwin S Dalmaijer, MRC Cognition and Brain Sciences Unit, 15 Chaucer Road, Cambridge, CB2 7EF, United Kingdom. edwin.dalmaijer@mrc-cbu.cam.ac.uk

**CRediT author contributions**

ESD, JvR, and JT conceptualised the metric. ESD wrote the software with input from EVS; collected and visualised the data; and drafted the article, which all authors reviewed and edited.

**Conflict of interest statement**

All authors declare to have no financial or non-financial relationships with any organisations (not even the Temperate Fruit Lobby) that influenced this work.

**Acknowledgements**

The keyword "banana" was chosen to mimic the popular internet meme "banana for scale", not because the authors think it is an inherently uninteresting scientific topic. (Please don't come after us on Twitter, banana research community!) This manuscript was submitted to the BMJ for its Christmas Edition, where it was deemed "amusing and quirky", but ultimately of too low priority. The current version incorporates helpful comments from BMJ's reviewers.

**Open data and software**

All data and code used for this manuscript can be found on github.com/esdalmaijer/bibliobanana

## Abstract

Many academics use yearly publication numbers to quantify academic interest for their research topic. While such visualisations are ubiquitous in grant applications, manuscript introductions, and review articles, they fail to account for the rapid growth in scientific publications. As a result, any search term will likely show an increase in supposed "academic interest". One proposed solution is to normalise yearly publication rates by field size, but this is arduous and difficult. Here, we propose an simpler index that normalises keywords of interest by a ubiquitous and innocuous keyword, such as "banana". Alternatively, one could opt for field-specific keywords or hierarchical structures (e.g. PubMed's Medical Subject Headings, MeSH) to compute "interest market share". Using this approach, we uncovered plausible trends in academic interest in examples from the medical literature. In neuroimaging, we found that not the supplementary motor area (as was previously claimed), but the prefrontal cortex is the most interesting part of the brain. In cancer research, we found a contemporary preference for cancers with high prevalence and clinical severity, and notable declines in interest for more treatable or likely benign neoplasms. Finally, we found that interest in respiratory viral infections spiked when strains showed potential for pandemic involvement, with SARS-CoV-2 and the COVID-19 pandemic being the most extreme example. In sum, the time is ripe for a quick and easy method to quantify trends in academic interest for anecdotal purposes. We provide such a method, along with software for researchers looking to implement it in their own writing.

## Introduction

There is an increased interest in "increased interest" in academic publishing (Figure 1), possibly reflecting an increasing need for academics to justify their research topic as growing in prominence. However, many academics do not take into account baseline publication rates in their field when presenting such visualisations (see e.g. (1), nostra culpa). Because academic outputs are growing across the board, one should expect an increase in publications for the majority of research topics. It is thus likely that uncorrected visualisations like Figure 1 are not as noteworthy as they are frequently presented to be.

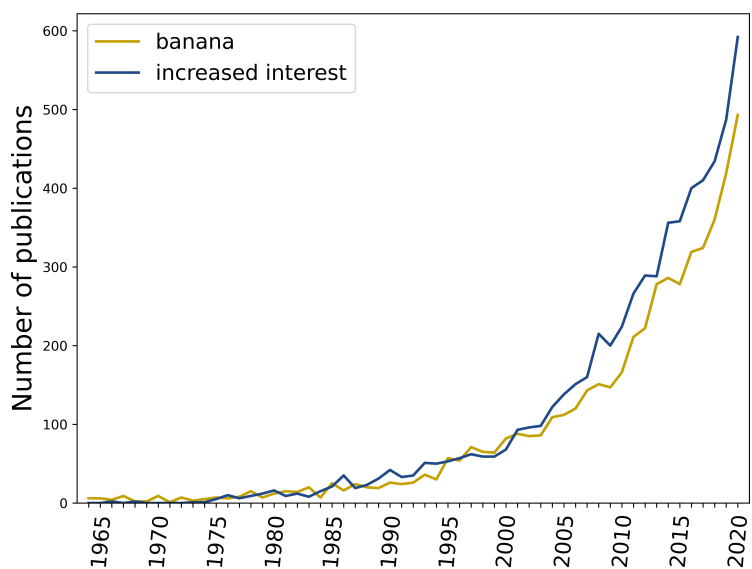

***Figure 1*** *– The yearly number of papers indexed in PubMed that reference "increased interest" (blue line) or "banana" (for scale). This trend was pointed our by Alex Bond on Twitter (https://twitter.com/TheLabAndField/status/1273598114708979713?s=20).*

      A similar problem exists in the field of academic impact measurement, particularly between fields with different publication and citation rates. Various methods have been proposed to account for this by normalising to field-specific rates (2–4). Unfortunately, not all bibliometric databases are as complete as they perhaps ought to be: Web of Science and Scopus do not cover all fields equally, are slow to incorporate new publications, and omit preprints. These are crucial roadblocks in the context of quantifying recent interest in a research topic. While databases like Google Scholar provide a more complete alternative, collecting and validating normalisation data through such databases requires extensive manual work (5).

      We have thus landed on a trade-off between accuracy and effort. To academics looking for a simple illustration that justifies their specific topic, proper normalisation techniques that accurately gauge changes in academic interest require more work than they are worth. The low-hanging fruit in armchair bibliography is thus a more user-friendly method for publication rate normalisation. Here, we present a novel metric of academic interest that is both easy to implement and "good enough"

for casual use. Specifically, we propose a normalisation of publications that mention a keyword of interest by those that mention a keyword of non-interest, such as "banana". We illustrate this approach in the fields of neuroimaging, cancer, and respiratory virology. Each of these show fluctuations in interest, sometimes particularly rapidly in the case of respiratory illnesses with pandemic potential (e.g. influenza A H1N1 and SARS-CoV-2).

## Methods

### Keyword normalisation

We quantified academic interest as the yearly number of publications that included a given keyword, divided by the yearly number of publications that included one (Equation 1) or several (Equation 2) comparison keywords. In Equation 1, this idea is demonstrated using "banana" as a reference keyword. In Equation 2, we expanded upon this idea by using the average of a set of reference keywords.

(1) $\quad e_{normalised} = \dfrac{p_{keyword}}{p_{banana}}$

(2) $\quad e_{normalised} = \dfrac{p_{keyword}}{\dfrac{1}{n} \sum\limits_{i=1}^{n} p_{comparison,i}}$

Where $e_{normalised}$ is the yearly academic interest (named $e$ for enthusiasm, and also for the first name of two of the authors), $p_{keyword}$ is the annual number of publications that reference a keyword of interest, $p_{comparison,i}$ is the annual number of publications that reference a comparison keyword $i$, and $n$ is the number of keywords in the comparison set.

### Keyword selection

We used "banana" as a comparison keyword because, in addition to being nutritious and versatile (6), it is both common and innocuous (although see (7)). These are ideal characteristics, as rare keywords could show unstable fluctuations, and more exciting keywords could show upward trends that exceed the general publication growth due to their inherent interestingness.

Bananas have received stable interest as a function of research that mentions "fruit" (Figure 2). More popular fruits were apples and oranges, which are inherently difficult to compare, and particularly so because "orange" could also appear as colour reference. This could explain why more published papers referenced "orange" than "fruit" in the late 1980s (8). Out of all fruits included here, "banana" showed the most stable publication ratio, driven by an increase in publications (12 in 1980, 26 in 1990, 83 in 2000, 170 in 2010, and 420 in 2019) that mirrored the general increase in scientific publications on fruit.

One fundamental limitation on keyword selection, is that no "neutral" term is without its caveats. For example, the publication of the banana genome in 2012 (9) could have inspired increased interest in banana research (perhaps because of its excellent use of Venn diagrams), although this has not yet become apparent in publication numbers (Figure 2).

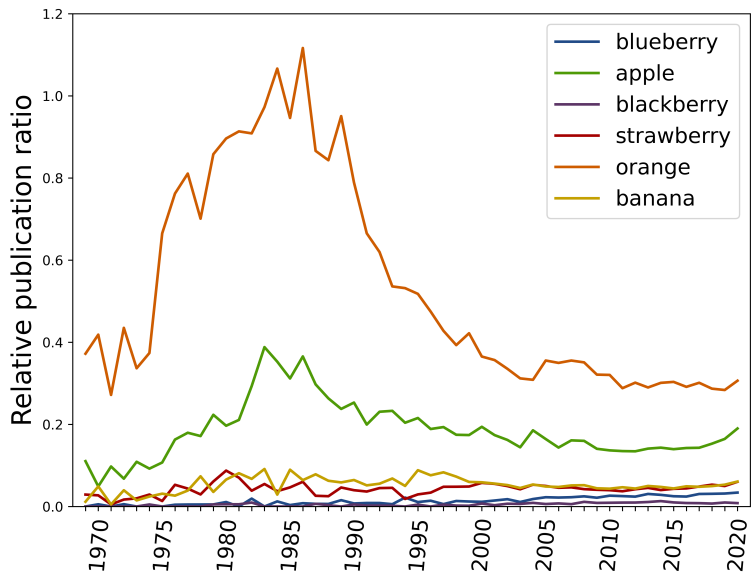

***Figure 2*** *– The yearly publication ratio (compared to keyword "fruit") of keywords "blueberry" (blue), "apple" (green), "blackberry" (purple), "strawberry" (red), "orange" (orange), and "banana" (yellow). This illustrates that apples and oranges cannot be compared: publications that reference "orange" far outnumber those that reference "apple". More importantly, "banana" shows numerous yet stable publication numbers, thus providing an excellent comparison keyword.*

As an alternative to "banana", field-specific comparison keywords could be chosen to account for differential publication growths between fields. We believe that semi-informal environments like Twitter or the doctors' mess (10) could play an important role in the process of establishing consensus between experts on their own fields' "banana" equivalents.

A more systematic way of choosing keywords, would be to tap into PubMed's Medical Subject Headings (MeSH), which have been used to track research trends (11). MeSH provides a hierarchical ordering for pathologies, and can thus be used to select reference keywords higher in the hierarchy for keywords of interest on lower rungs.

**Example keywords**

To provide an example of a field-specific keyword, we examined trends in the neuroimaging literature by comparing several brain areas to the keyword "brain". This keyword was chosen because it is both ubiquitous in neuroscience, and relatively unbiased between brain areas. The increase in publications that mention "brain" is thus expected to reflect the general increase in human neuroimaging research output.

Another example could be the reference keyword "cancer" to normalise research on specific sub-types. This comparison computed the proportion of work within the wider field of cancer research, and thus reveals the relative popularity of specific cancers. When used in this way, our suggested keyword-normalisation approach can be used to reflect the "market share" for each search term in a subset that contains the reference keyword.

The third example included here is that of influenza and COVID-19, which are expected to show periodic trends in interest that align with outbreaks of specific strains in humans. We employed reference keyword "virus", expecting it to reflect trends in the wider field.

Finally, to demonstrate keyword selection from the MeSH tree, we investigated relative publication trends under the MeSH term "Endocrine Gland Neoplasms" (C04.588.322), which includes nodes "Adrenal Gland Neoplasms" (C04.588.322.078), "Ovarian Neoplasms" (C04.588.322.455), "Pancreatic Neoplasms" (C04.588.322.475), "Pituitary Neoplasms" (C04.588.322.609), "Testicular Neoplasms" (C04.588.322.762), and "Thyroid Neoplasms" (C04.588.322.894). These keywords capture articles relating to specific subtypes of endocrine gland tumours, and will act as an example for comparing publication ratios between terms within the MeSH hierarchy.

**Search queries**

Searches were performed using the NCBI E-utilities API (12) (see below for details). Queries were performed per year, in the PubMed database, using the "TEXT" field for keywords in examples 1-3 or the "MESH" field in example 4, with double-quoted terms. For example, a search for a specific influenza strain in the year 2018 could be ""H1N1"[text]+AND+2018[pdat]", with "db=pubmed" and "rettype=count". Data for the current manuscript were collected on 8 February 2021.

**Open software and data**

We developed a Python library to count and normalise the number of publications that reference any keyword. It interfaces with the NCBI E-utilities API (12) to count publication numbers on PubMed. Similar interfaces exist as online tool PubMed By Year (https://esperr.github.io/pubmed-by-year/), and as R package (13) to interface with Europe PMC (14).

The source code can be freely accessed on GitHub, and the package can also be downloaded from the Python Package Index (PyPI). Example scripts are included, as are the data analysed here. See: https://github.com/esdalmaijer/bibliobanana

**Statistical tests**

Due to the anecdotal purpose of the proposed index, and to avoid encouraging the misappropriation of null-hypothesis significance testing, no statistical tests were performed.

## Results

### Example 1: Publication trends in neuroimaging reveal the most interesting brain area

We examined the period from 1990 until 2020. We compared a selection of brain areas: prefrontal cortex, visual cortex, cingulate cortex, amygdala, supplementary motor area, and frontal eye fields.

Since 1990, the number of publications referencing brain areas has sharply increased (Figure 3A). While the raw number of publications indicated an upwards trend for all included brain areas, the banana ratio (Figure 3B) revealed that academic interest was in fact stable or declining for all but the prefrontal and cingulate cortices. (An alternative conclusion could be that interest in banana research has grown faster than interest in the brain.) Importantly, normalising against field-specific keyword "brain" (Figure 3C) confirmed that academic interest steadily decreased for the visual cortex and amygdala; remained constant for supplementary motor area and frontal eye fields; while interest for the prefrontal cortex and cingulate cortex increased.

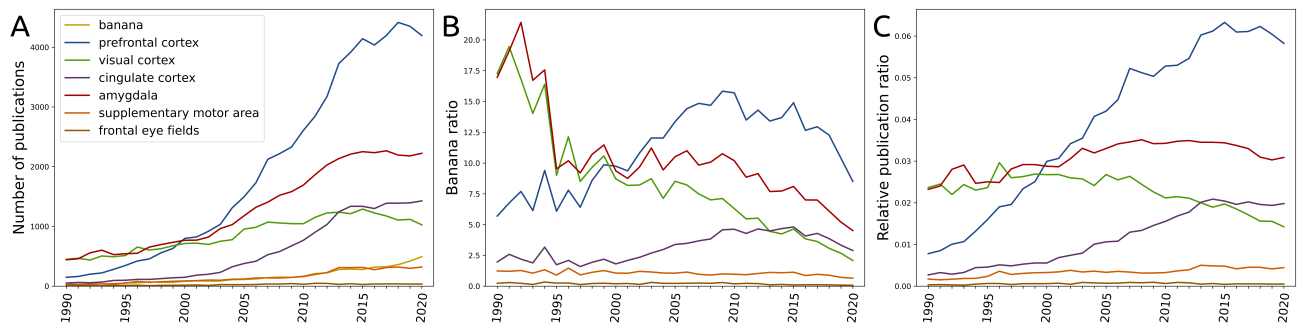

***Figure 3*** *– A) The number of publications that mention "prefrontal cortex" (blue), "visual cortex" (green), "cingulate cortex" (purple), "amygdala" (red), "supplementary motor area" (orange), "frontal eye fields" (brown), or the word "banana" (yellow) increased every year. B) Relative trends are more apparent in the banana ratio. C) Field-specific trends are further clarified after scaling the number of publications for each brain area by common and innocuous keyword "brain", revealing that prefrontal cortex and cingulate cortex are increasing in relative popularity, while visual cortex and amygdala are decreasing.*

It has previously been argued that the supplementary motor area is the most interesting part of the brain (15), but these results indicate that this view is not supported by the wider neuroimaging literature. Yearly publication numbers suggest that the supplementary motor area is almost exactly as interesting as a banana. Instead, continuous growth in relative interest for the prefrontal cortex indicates that it is not only in the front of researchers' brains, but also in the front of their minds.

**Example 2: Publication trends in cancer research "market shares"**

As a second example from the medical literature, yearly numbers of publications about different types of cancer were normalised against the broad keyword "cancer".

This analysis (Figure 4A) revealed that interest in skin cancer has remained relatively stable since 1970; whereas interest has increased gradually for breast, lung, and colorectal cancer, and more rapidly for prostate cancer from the 1990s. Notably, relative interest in testicular cancer has decreased since 1985. This coincides with the introduction (16) and proven superiority over existing treatments (17) of bleomycin, etoposide, and cisplatin; the current gold-standard treatment that has greatly reduced mortality since its introduction (18). One could speculate that this reflects a lack of interest in "solved" diseases among the medical research community.

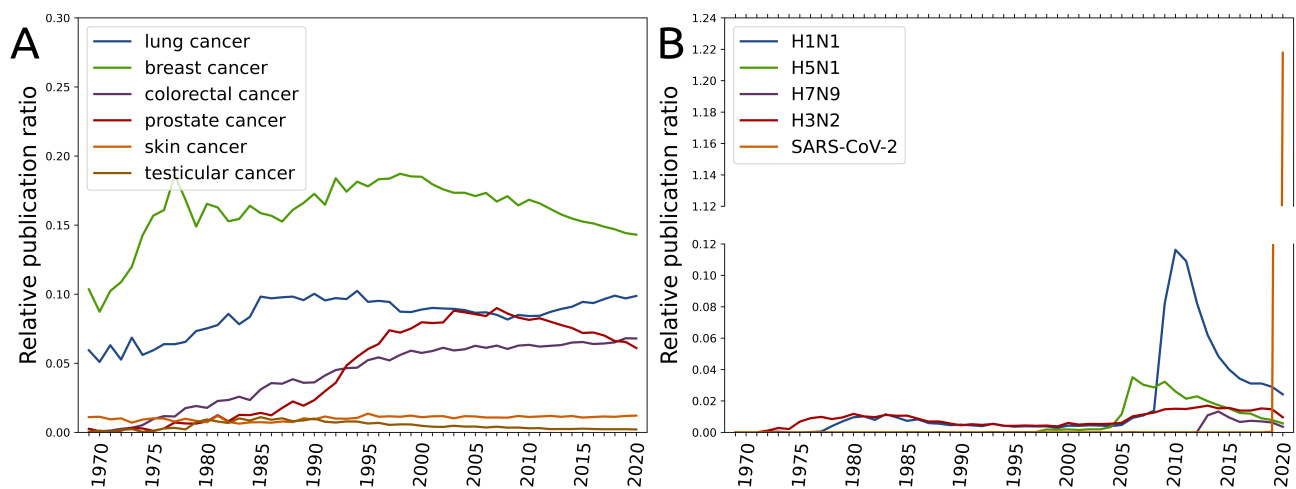

***Figure 4*** *– A) The yearly number of publications that mention specific types of cancer (lung in blue, breast in green, colorectal in purple, prostate in red, skin in orange, testicular in brown), normalised by the number of publications that mention "cancer" to reveal relative interest. B) The number of publications that mention specific influenza strains (H1N1 in blue, H5N1 in green, H7N9 in purple, H3N2 in red) or SARS-CoV-2 (orange), normalised by the number of publications that contain keyword "virus". Note that the y-axis is discontinuous, and jumps from 0.12 to 1.12.*

**Example 3: Publication trends in respiratory disease mirror potential outbreaks**

One field with clear and acute trends, and thus an excellent test-case for our bibliographic interest method, is that of viruses that cause respiratory illness. We compared several strains of influenza and SARS-CoV-2 with the common keyword "virus". This keyword was chosen to account for the technological advances that support the viral increase in publication output.

This analysis (Figure 4B) confirmed our metric's ability to identify periodic trends amid generally increasing publication numbers. Notably, relative interest for each influenza strain peaked after potential pandemic involvement: H3N2 in 1968 (19), H5N1 after a public warning in 2003 (20), H1N1 in 2009 (21), and H7N9 after the first human infections were reported in 2013 (22).

Interest in the virus SARS-CoV-2 has quickly grown to dwarf that for influenza since its emergence in 2019. At the time of data collection (8 February 2021), more papers were published that mentioned SARS-CoV-2 than there were papers that mentioned reference keyword "virus". This highlights the importance of choosing keywords that encompass the field of interest.

**Example 4: Publication trends under MeSH term "Endocrine Gland Neoplasms"**

This final example served to illustrate relative differences between terms in the MeSH hierarchy. Here, six terms captured under "Endocrine Gland Neoplasms" were compared.

This analysis showed that the number of publications relating to endocrine gland tumours increased rapidly in the last decades (Figure 5A), and highlighted trends in interest (Figure 5B). Specifically, academic interest has been consistently high for ovarian neoplasms; with pancreatic neoplasms overtaking the top position in 2012, after a steady climb in publication rate. A more modest increase was apparent for thyroid tumours, whereas it dropped for other neoplasms.

The decrease in publication rates for testicular neoplasms from 1985 was also apparent under Example 2, where it was attributed to the introduction of a new gold-standard treatment that rendered testicular cancer much more treatable. In this example, highly similar reductions in interest are apparent for tumours of the pituitary and adrenal glands. Both of these neoplasms are characterised by their rarity and high likelihood of being benign. Hence, the decrease in relative popularity for testicular, pituitary, and adrenal neoplasms could indeed reflect a trend away from conditions with lower clinical impact.

A large dip in publication numbers is apparent in 2020. This is artificial, and due to a delay in MeSH indexing after articles are entered into PubMed (23).

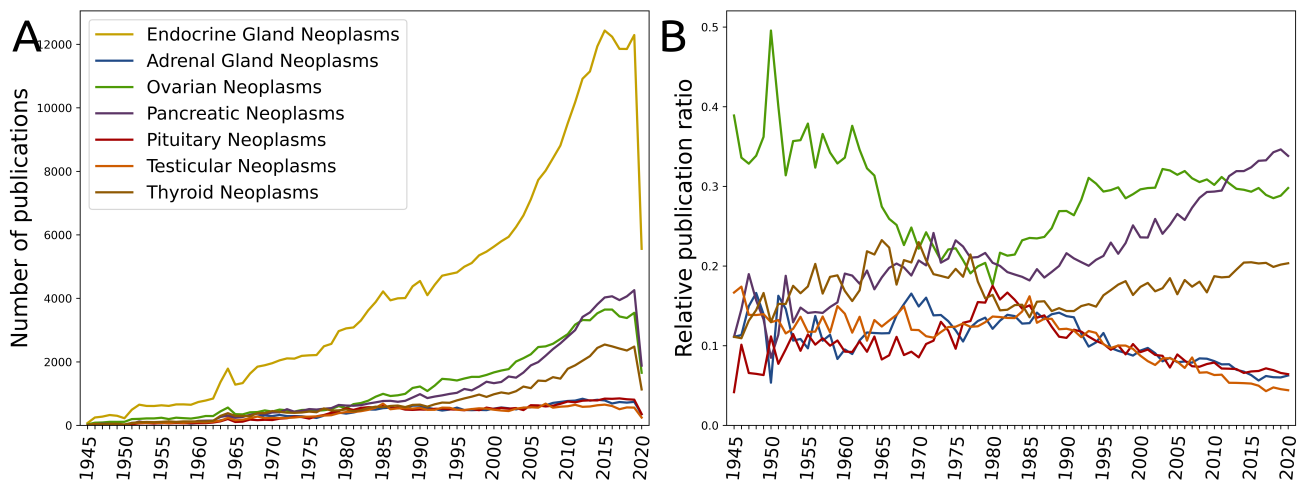

***Figure 5*** – *A) The yearly number of publications in PubMed MeSH terms "Adrenal Gland Neoplasms" (blue), "Ovarian Neoplasms" (green), "Pancreatic Neoplasms" (purple), "Pituitary Neoplasms" (red), "Testicular Neoplasms" (orange), and "Thyroid Neoplasms" (brown); which are members of the MeSH term "Endocrine Gland Neoplasms" (yellow). B) The relative ratios for sub-terms of the "Endocrine Gland Neoplasms" MeSH term.*

## Discussion

While there is an abundance of complex normalisation tools to put publication and citation rates into perspective, there are few simple-yet-accurate alternatives for researchers who wish to quantify academic interest for their grant proposals or manuscripts' introduction sections. We proposed a metric that compares the ratio of a keyword of interest to one or several 'steady interest' keywords, allowing for quick normalisation of bibliometric trends against general increases in publications within a field of research. We validated this metric in the fields of neuroimaging, cancer, and respiratory illnesses, using both general ("banana") and field-specific keywords. This revealed that while general publication rates increased dramatically over the last decades, the "banana ratio" only increased for a subset of truly interesting topics, and normalising against field-specific keywords further clarified these trends.

In conclusion, academics can use our metric (and freely available software) to easily compute and visualise bibliometric trends. These visualisations can be used as informal but accurate illustrations of how popular or overlooked the topic of their next paper or grant is.